Title: **Biological Effects of Gamma-Ray Bursts: distances for severe damage on the biota**

Authors:

Douglas Galante

Instituto de Astronomia, Geofísica e Ciências Atmosféricas, Universidade de São Paulo

Jorge Ernesto Horvath

Instituto de Astronomia, Geofísica e Ciências Atmosféricas, Universidade de São Paulo

Correspondence:

Douglas Galante

Rua do Matão, 1226, Cidade Universitária, 05508-900 São Paulo – SP, Brazil

Phone: (+5511) 3091-2736

Fax: (+5511) 3091-2860

Email: douglas@astro.iag.usp.br

**Abstract**

We present in this work a unified, quantitative synthesis of analytical and numerical calculations of the effects that could be caused on Earth by a Gamma-Ray Burst (GRB), considering atmospheric and biological implications. The main effects of the illumination by a GRB are classified in four distinct ones and analyzed separately, namely: *direct γ Flash, UV Flash, Ozone Layer Depletion and Cosmic Rays*. The effectiveness of each of these effects is compared and distances for significant biological damage are given for each one. We find that the first three effects have potential to cause global environmental changes and biospheric damages, even if the source is located at galactic distances or even farther (up to 150 *kpc*, where $1 pc = 3.09 \times 10^{16} m$, about five times the Galactic diameter of 30 *kpc*). Instead, cosmic rays would only be a serious threat for close sources (on the order of a few *pc*).

As a concrete application from a well-recorded event, the effects on the biosphere of an event identical to the giant flare of SGR1806-20 on Dec 27, 2004 have been calculated. In spite of not belonging to the so-called "classical" GRBs, most of the parameters of this recent flare are quite well-known and have been used as a calibration for our study. We find that a giant flare impinging on Earth is not a threat for life in all practical situations, mainly because it is not as energetic, in spite of being much more frequent than GRBs, unless the source happens to be extremely close.

**Introduction**

The implications of astrophysical events on Earth and Earth-like planets is an increasingly active field of research in many and distinct areas of knowledge, such as Astronomy, Geology, Meteorology and Biology. Several sources of cosmic catastrophes have been proposed over the years. Recently, Gamma-Ray Bursts were recognized as some of the most energetic astrophysical events since the Big Bang, releasing in few seconds as much energy as a supernova does, of the order of $10^{44}\,J$ but concentrated in hard X-rays and $\gamma$ radiation. Thorsett (1995) was the first to acknowledge the potential destructive effects of a GRB illuminating the Earth. The question of vital statistics (rates, beaming, etc.) of the bursts is nowadays being studied, but it is undeniable that damage to the biota could be severe if a burst strikes (or has struck) the planet. Therefore, it is important to understand the various dangerous effects, and especially their relative importance as a function of the distance for realistic physical inputs, in order to assess the actual threat to living organisms. It is the purpose of this work to present a unified synthesis of numerical and analytical calculations on the atmospheric and biological effects a GRB might have if directed to Earth or to a planet with an Earth-like atmosphere.

It is not unlikely that GRBs (as well as SNe) have had great impact on Earth, at least on the last billion years (Melott *et al.* 2004). Their effects range from direct transmission of the high energy $\gamma$ radiation through the atmosphere to chemical alterations on it, such as $NO_x$ rise and ozone layer depletion. GRBs may also be associated (Vietri *et al.* 2003) with the acceleration of high energy cosmic rays, which were taken on account by considering the effects of a massive flux of particles, on the form of a Cosmic Ray Jet (CRJ), as proposed by Dar *et al.* (1998).

As demonstrated below, our results suggest that "classical" GRBs can have a radius of biological influence as high as of 150 *kpc*. This distance is greater than the galactic radius, and therefore a burst happening anywhere in the Galaxy could directly affect the biosphere of a planet lying within its beaming cone. It is naively expected that galaxies with strong star formation should be more prone to the occurrence of GRBs, due to the greater proportion of massive stars, which are considered to be the progenitors of long duration bursts in most of the models. If life was once originated there, it might have been "reset" by a GRB (Annis 1999), or even it may have received an evolutionary boost. Irregular galaxies with low metallicity may also be associated with GRBs (Stanek 2006), but these systems are less interesting for astrobiology, because it is still unclear if planets could be formed on them at all.

1. **Basic assumptions**

The model adopted for a long duration GRB consists of a "standard" γ energy release of $5 \times 10^{43} J$, beamed within a solid angle $\Delta\Omega \sim 0.01$ sr (Frail *et al.* 2001) and peak duration of ~ 10s. The emission is generally well-fit by a broken power-law, known as *Band spectrum*, with its main energy being released on the hundreds of *keV* range (Band *et al.* 1993), where $1 eV = 1.6 \times 10^{-19} J$. The absorption by the interstellar medium (ISM) was neglected because the later is almost completely transparent to high energy photons, thus, the flux at a distance $D$ should behave as $F = \frac{L}{\Delta\Omega D^2}$, where $L$ is the γ luminosity and $\Delta\Omega$ is the beaming solid angle.

Since we shall be using the observational characteristics of long bursts only, we will not take on account the differences of the progenitors which may lead to the event itself, nor the different models to explain the formation of the γ radiation, being the Fireball and the Cannonball models the most popular ones. The Fireball model associates this kind of burst with very energetic Ic SNe, dubbed sometimes *hypernovae* events (van Paradijs *et al.* 2000). In this model the γ radiation is produced by synchrotron emission during the collision of highly relativistic conical shells ejected during the explosion. In the Cannonball model (Dar and De Rújula 2004), the γ emission is produced by inverse Compton scattering on a bipolar jet made of chunks of ordinary material from a common core-collapsed supernovae, which is ejected when material from an accreting torus falls down on the compact object. In spite of being important for other purposes, these differences do not affect our conclusions, which are insensitive to the specific GRB model.

Some works have already made quantitative estimates on individual effects, for example, Thomas *et al.* (2005) made a study of the ozone depletion produced by a GRB using several initial conditions. Since we seek here to build a more complete picture of the problem, we have also considered the effects of transmission of the γ radiation by the atmosphere (Smith, Scalo and Wheeler 2004) and the energy deposited by cosmic rays accelerated within the burst. Thus, for the sake of a clear identification of the different harmful thresholds for the biota related to each effect, we have divided them into four classes: γ Flash, UV Flash, $O_3$ Layer Depletion and CRJs, which will be further discussed and quantified.

After an evaluation of the physical conditions of the photon transfer, two unicellular organisms were used as fiducial biological probes of their effects. The well-known *Escherichia coli*, an internal, radiation-sensitive (specially UV) bacterium, and *Deinococcus radiodurans*, classified as a polyextremophyle due to its resistance to many external agents, such as UV and ionizing radiation, organic peroxides and desiccation. They represent two extremes on radiation resistance, defining a pragmatic "surviving zone for life". We are not claiming these organisms to be representative of the primordial prokaryotic fauna on Earth, but they happen to be useful as biological standards because of the vast amount of available data on their biology. Their radiation resistances are summarized in TABLE 1 and a discussion on the population dynamics is further developed on Appendix A.

**TABLE 1**

## 2. Evaluation of $D_{10}$ distances

### a. Direct γ Flash

As discussed, for instance, in the recent numerical simulations by Smith, Scalo and Wheeler (2004), γ photons with energies of the order of hundreds of *keV* loose energy primarily by Compton scattering over the electrons on the atmosphere. During the transmission through the atmosphere, these photons loose energy, when characteristic energies decrease from around 250 *keV* to 20 *keV*.

Following Smith, Scalo and Wheeler's calculations, a thin atmosphere, such as the one present on primordial Earth, with $\rho < 100$ *g/cm²* would transmit around 1% of the initial γ radiation to the ground, while the present atmosphere, a much thicker one ($\rho \sim 1024$ *g/cm²*), would transmit only a fraction of around $6 \times 10^{-29}$ of the initial γ fluence. This means practically all the γ energy of the burst would be deposited on a thick atmosphere, while on a thin one, a significant fraction of the radiation would still reach the ground, affecting just the illuminated hemisphere, because of the short duration of the burst (typically around 10 *s*), certainly much shorter than the rotation of the Earth.

The main biological effect of γ radiation of this range of energy is free radicals production by ionization, which have toxic effects on the cell. The $D_{10}^{ion}$ dose for *E.coli* is $D_{10}^{ion} = 0.7 kGy$ where $D_{10}^{ion}$ stands for the necessary dose of ionizing radiation for a 10% survival. For *D.radiodurans*, the $D_{10}^{ion}$ is much higher, $D_{10}^{ion} = 11 kGy$. Theses doses can be converted to fluxes by using the energy losses on water. For *E.coli*, we have $F_{10}^{ion} = 3.5 \times 10^5 J/m^2$ and for *D.radiodurans* $F_{10}^{ion} = 5.5 \times 10^6 J/m^2$ of γ flux. In order to get such fluxes on the surface of the planet, we have to consider the transparency of the atmosphere to the γ radiation, as stated above.

The maximum distance at which the γ flash would deliver a $D_{10}^{ion}$ dose at the surface of the planet was calculated to be, for *E.coli*, with a thin atmosphere, of 390 *pc*, and for *D.radiodurans* of 100 *pc*. For the thick atmosphere, this mechanism would be very inefficient as stated, since the γ radiation would be deposited on the atmosphere itself.

**b. UV Flash**

The same Compton mechanism blocking the direct irradiation provokes a large fraction of the high energy γ photons to have its energy lowered to the UV range.

Smith, Scalo and Wheeler (2004) calculated that, for a thin atmosphere, 1 to 10% of the initial γ flux would be converted to UV flux, while for a thick atmosphere the fraction is in the range $2 \times 10^{-3}$ to $4 \times 10^{-2}$. We observe that these calculations were performed without considering the presence of aerosols or clouds, which could quite effectively block the UV radiation. As this effect would last for just a few seconds, this means that the instantaneous condition of the atmosphere might change substantially the amount of radiation arriving on the ground.

The biological effectiveness of the UV radiation as a harmful factor comes from the fact that DNA and RNA strongly absorb in this range of energy, suffering mainly from nucleotide dimerization, especially on the pirimidines (Häder and Sinha 2005). In fact, the UV flux necessary to deposit a $D_{10}^{UV}$ dose on *E.coli* is $F_{10}^{UV} = 22.6 J/m^2$ (Gascón et al. 1995), a value $10^4$ times lower than from ionizing radiation. Again, for *D.radiodurans*, this value is higher, $F_{10}^{UV} = 5.53 \times 10^2 \, J/m^2$ (Ghosal et al. 2005). These fluxes can be also translated into distances. Using the same procedure as before and considering the γ to UV efficiencies given by Smith, Scalo and Wheeler (2004) and summarized above, we found that the γ flux on the top of the atmosphere has to be 10 - 100 times greater than the $F_{10}^{UV}$ on the ground, which implies on a maximum distance to deposit such doses on the range of 48 *kpc* to 152 *kpc* for thin atmospheres for *E.coli* and 10 *kpc* to 31 *kpc* for *D.radiodurans*. For the thick atmosphere case, the range is of 21 *kpc* to 96 *kpc* for *E.coli* and 4 *kpc* to 9 *kpc* to *D.radiodurans*. These ranges of distances arise from the uncertainty on the γ to UV conversion, as presented above.

### c. *$O_3$ layer depletion*

The γ radiation may alter the chemistry of the atmosphere, most interesting for our case being the rise of the concentration of *$NO_x$*, which can act as catalyzers for *$O_3$* degradation. This problem has already been addressed in the 70s (Ruderman 1979) related to the effects of SNe explosions on the atmosphere.

Ruderman's calculations are based on the catalytic destruction of *$O_3$* by *NO*, as follows

$$NO + O_3 \rightarrow NO_2 + O_2$$
$$NO_2 + O \rightarrow NO + O_2 \quad (1)$$

The kinematics of the reaction was modeled in a simplified manner by the equation

$$\frac{[O_3]}{[O_3]_0} = \frac{\sqrt{19 + 9X^2} - 3X}{2} \quad (2)$$

$$X = \frac{[NO_X]}{[NO_X]_0} \tag{3}$$

Further assumptions of Ruderman's work included:

- The oxidation of $NO_2$ by $O$ is the limiting step of the cycle;
- The $[O]/[O_3]$ ratio is taken to be constant.

During the γ irradiation, the *[NOₓ]* rises due to the production of free N atoms

$$N^* + O_2 \rightarrow NO + O \tag{4}$$

If we consider all the $NO_x$ to be *NO*, the production rate is given by

$$\frac{10}{(10+y)} = \frac{NO \text{ molecules}}{\text{ion pair}} \tag{5}$$

where *y* is the initial concentration in *ppb*, which will be taken to be 3 *ppb*.

Using Ruderman's figures (Ruderman 1979), we have that 1 *J* of γ radiation produces $2.8 \times 10^{17}$ ion pairs. The rate of production of NO molecules is then

$$R_{NO} = \phi(J/m^2/s) \times 2.8 \times 10^{17} \left(\frac{\text{ion pairs}}{J}\right) \times$$

$$\times \left(\frac{10}{10+y}\right)\left(\frac{NO \text{ molecules}}{\text{ion pairs}}\right) \times \frac{10^9 \, (ppb)}{5 \times 10^{27} \, (\text{molec/m}^2)} \tag{6}$$

Simplifying and integrating over the duration of the burst yields

$$R_{NO} = 0.43\phi(J/m^2) \, ppb \tag{7}$$

Using the definition (3) with *[NO]₀* = 3 *ppb*, we finally obtain

$$X = \frac{[NO]}{[NO]_0} = 1 + \frac{R_{NO}}{[NO]_0} \tag{8}$$

$$X(\phi(J/m^2)) = 1 + 0.67 \times 10^{-2} \phi(J/m^2) \tag{9}$$

The *NO* thus produced has a dynamical time of residence in the atmosphere of the order of 2-6 years (Ruderman, 1979). Using equation (1), we may obtain the ozone depletion in function of the γ fluence on the top of the atmosphere, as presented on FIG 1. Ellis and Schramm (1995) later considered the same problem, with the addition of the effect of charged particles produced by the SN event, considering its ionizing power to be the same as the electromagnetic component. Their results are even more dramatic and may be adapted immediately for a GRB.

Recently, Gehrels and collaborators (Gehrels *et al.* 2003) did a more sophisticated treatment for the problem, using a 2D atmospheric simulation code developed at Goddard Space Flight Center, and using an input spectrum from SN1987a, with total energy release of $9 \times 10^{39} J$. Running the simulation at 10, 20, 50 e 100 *pc*, their results suggest a tendency of ozone depletion scaling as $D_{SN}^{-n}$, where $1.3 < n < 1.9$. We used for our calculations the mean value of $n = 1.6$.

We calculated the mean ozone depletion at 10 *pc* using the latest results from Gehrels *et al.* (2003) to be 22.6%, and from this value, we scaled to the fluences of interest, obtaining the depletion of $O_3$:

$$\frac{[O_3]}{[O_3]_0} = 1 - 7.2 \times 10^{-4.6} \phi^{0.8} (J/m^2) \tag{10}$$

**FIG 1**

Comparing Ruderman's results with Gehrels *et al.*, we appreciate the large differences between both approximations. Ruderman's results indicate that with fluences of $10^3 \, J/m^2$ essentially all the $O_3$ would be destroyed, while the recent simulations show that a much modest fraction ensues. This is caused by the strong simplifications adopted in Ruderman's work, especially on the chemical reactions and by neglecting atmospheric mixing, so that the problem could be solved analytically (Crutzen and Brühl 1996). We have thus adopted the milder scenario presented by Geherels *et al.* (2003). In particular, Thomas *et al.* (2005) repeated the calculations using the same code, but now with an input spectrum of a GRB, which has a much shorter time scale than a SN. However, their results did not change significantly our conclusions.

To calculate the increase of the solar UV caused by the depletion of the $O_3$ layer, we used a Beer-Lambert law

$$\phi_{UV} = (\phi_{UV})_0 e^{-\sigma N} \quad (11)$$

Where $\sigma$ is the $O_3$ cross section and $N$ is the column density of $O_3$. In this approximation we are ignoring other UV absorbers, such as water vapor and scattering particles, which are not expected to dominate. The initial mean value for the column density of $O_3$ used was 350DU, or equivalently $9.4 \times 10^{22} \, molec/m^2$.

We may now introduce the depletion factor from equation (10) into equation (11), thus obtaining the solar $UV_B$ flux on the ground already increased by the depletion of the $O_3$ layer due to the $\gamma$ flux from the GRB:

$$\phi_{sup}(J/m^2/s) = \phi_0(J/m^2/s) e^{-\left(\sigma \frac{[O_3]}{[O_3]_0} N\right)} \quad (12)$$

**FIG 2**

The results are depicted in FIG 2. The $D_{10}^{UV}$ distance for D.radiodurans is of 12 kpc, essentially the Galactic radius. Closer bursts increase dramatically the mortality for almost all types of exposed unicellular organisms. The *E.coli* $F_{10}^{UV}$ is bellow the actual solar UV flux on the ground, even without ozone depletion, which is consistent with the fact of being an internal microorganism and does not result in a useful limit.

### d. Cosmic Ray Jets

This is the last considered effect in our work. Being an explosive phenomenon, Dar, Laor and Shaviv (1998) proposed that GRB may be associated with massive acceleration of cosmic rays, which could be beamed into a jet and reach great distances from the source, termed by them as *Cosmic Ray Jet* (CRJ). Although still very speculative, it is worth to take a serious look to this hypothesis, especially on the consequences it might have to a planet illuminated by the GRB, since it could also be struck by the jet, but with a substantial time delay.

It is generally assumed in the calculations that the same amount of energy seen in γ radiation is used to accelerate the particles (termed the equipartition hypothesis, see for example, Vietri *et al.*, 2003), which we shall consider to be protons. This way, we considered a jet of protons with $5 \times 10^{43} J$ of kinetic energy and assumed the same collimation angle inferred for the photons, $\Delta\Omega = 1/\Gamma$, with $\Gamma \approx 100$.

Hitting the upper atmosphere, the CRJ would produce a shower of secondary particles. We focused our attention on the muons, which could arrive on the ground and even underground or deep underwater, unlike UV or γ radiation that are restricted to the surface. In fact this is one of the main arguments given by Dar, Laor and Shaviv (1998) of why CRJs could probably have a major impact on life.

The production of muons on the atmosphere occurs when a proton interacts with a nucleus in the sequence:

$$p + N \rightarrow \pi + K \rightarrow \mu + \nu \qquad (13)$$

In the case of a monochromatic primary flux, we can use the simplified formula (Dar, Laor and Shaviv 1998)

$$\langle N_\mu \rangle \cong \frac{(0.0145 E_p [TeV])}{\cos\theta} \left( E_p / E_\mu \right)^{0.757} \left( 1 - E_\mu / E_p \right)^{5.25} \qquad (14)$$

This gives the mean number of high energy muons (E > $E_\mu$) produced by nucleons of energy $E_p$, which do not decay in the atmosphere and reach the sea level with zenithal angle $\theta < \pi/2$.

To introduce the incident spectrum of the primaries, we can use the equation given by Lipari (1993), deducible using standard physics (eg, Gaisser, 1992):

$$\phi_\mu(E,\theta) = \left[ L_\pi(\alpha) \left( 1 + \frac{L_\pi(\alpha)}{H_\pi(\alpha)} \frac{E\cos\theta}{\varepsilon_\pi} \right)^{-1} + L_K(\alpha) \left( 1 + \frac{L_K(\alpha)}{H_K(\alpha)} \frac{E\cos\theta}{\varepsilon_K} \right)^{-1} \right] K E^{-\alpha} \qquad (15)$$

These formulae are strictly valid for muon energies greater than 20 *GeV*, because muon decay processes are not taken into account. The input spectrum is characterized by the index α and the constant K, these parameters were taken from the work of Lipari (1993). This formula was tested with success to the usual cosmic ray background ($\alpha$=2.7 and K=1.85) reproducing the actual flux, except at low energies where the solar wind and other effects are important.

The main process of energy loss of high energy muons on matter is ionization. To calculate the ionization losses, we used a simplified well-known version of the Bethe-Bloch equation (Richard-Serre, 1971), which is valid strictly for energies above 10 *GeV*:

$$\left(\frac{dE}{dx}\right)_{ioniz} = \frac{Z}{A}\left[4.55 + 2.34\times 10^{-4}\eta - 2.62\times 10^{-8}\eta^2 - 0.1535\ln\left(\rho\frac{Z}{A}\right)\right] \quad (16)$$

$$\eta = \beta\gamma = \left[\left(\frac{E}{M_\mu c^2} - 1\right)\frac{E}{M_\mu c^2}\right]^{1/2}$$

In water, equation (16) gives an approximate constant energy loss of 2.58 *MeV/g/cm²*, for muon energies up to 1 *TeV*.

Lacking of a firm evaluation of the spectrum, we have decided not to employ any specific one for the particles accelerated at the GRB, but rather a monochromatic flux of typical energy per nucleon of 1 *TeV*. The duration of the irradiation by these relativistic particles is estimated to be $\approx$ 2 months (Dar *et al.* 1998).

For a primary energy flux of $10^4 J/m^2 \cong 10^{12} TeV/m^2$ in high-energy particles, which is the assumption for our standard burst at 10 *kpc*, we have calculated the muon flux at sea level shown if FIG 3

**FIG 3**

The lethal dose of ionizing radiation for humans is around 3 *Gy,* which can be translated to a muon flux, at the 20 *GeV* energy range, of about $10^{14}$ *m⁻²*. Our benchmark organisms *E.col*i and *D.radiodurans* are much more resistant, the $D_{10}^{ion}$ dose being 0.7 *kGy* for *E.coli* and 11 *kGy* for *D.radiodurans*, therefore, these bacteria could stand higher muon fluxes, $2\times10^{16}\,m^{-2}$ for *E.coli* and $3\times10^{17}\,m^{-2}$ for *D.radiodurans*. These numbers set $D_{10}^{ion}$ distances for the burst source of 300 *pc* for humans, 48 *pc* for *E.coli* and 12 *pc* for *D.radiodurans*.

### 3. Discussion and conclusions

As a summary of the above results, we present in TABLE 2 the $D_{10}$ distances referred to *E.coli* and *D.radiodurans* calculated for the various mechanisms presented.

**TABLE 2**

We can safely state that the most efficient damaging effect of GRB illumination is the UV flash, because it can be deliver a $D_{10}$ dose for distances up to 150 *kpc*. However, this effect is limited to one hemisphere, and only over uncovered land and shallow waters. It may not have a direct global impact, although it may have an indirect long term effect if a significant part of the planktonic organisms is killed during the irradiation. The non-linear effects on populations of a huge catastrophe like the incidence of a nearby GRB are difficult to model, and there is ample room to study scenarios addressing these issues.

The direct γ Flash seems not to be biologically important, because most of its energy would be absorbed by the atmosphere. For a thick atmosphere, the energy deposition would probably heat it up in a few degrees, but for a thin one, the results could be even more dramatic. However, the climatic consequences of these are not totally clear, and it would be interesting to model such a large disturbance in some detail.

The depletion of the ozone layer is the most obvious global and long lasting effect. It can affect life for many years, probably making the surface of the planet an environment not appropriate for its photosynthesizing biota. It can be effective for distances up to 12 *kpc* for *D.radiodurans*, which means almost anywhere in the Galaxy, even for a radiation resistant organism, confirming the expectations given by Thorsett (1995). In fact, it is difficult to envision a fundamental ecosystem depending on photosynthetic organisms not to be, at least, harmed by the occurrence of a directed GRB event closer than a few *kpc*. UV radiation has been proposed to have a role in extinctions on Earth (Cockell 1999), and it is important to consider GRBs as additional sources of ozone depletion, for their high efficiencies in doing so.

Other consequences are expected as well, as pointed by Thorsett (1995) and Thomas *et al.* (2005b): the rise of the $NO_x$ concentration on the atmosphere may have a global cooling effect, blocking visible sunlight and making photosynthesis inefficient. On the other hand, the residual nitrates of this process may make the soil more fertile after the end of this GRB winter, allowing lands to be populated by vegetation, as suggested by Thomas *et al.* (2005b).

The cosmic rays effect is still controversial, and it seems to be very inefficient unless the source is located very close (few pc) to Earth. However, a non-lethal CRJ incidence could still be important for the biota, for example by providing a higher level of background radiation which could induce significant mutation rates. Because its effect lasts for several months, these mutations might have time to accumulate on simple, fast replicating organisms, having a yet unknown evolutionary importance (Dermer and Holmes 2005).

We conclude our unified study of the several effects of GRBs with a quantitative assessment of how destructive an event could be, leading to extermination of life, or at least, part of it. However, it is not impossible that it may work as an evolutionary kick inducer, as many other apparent catastrophic events that happened on Earth (Horvath 2003).

### 3.1 Case study: SGR1806-20

On December 2004, a giant flare event from the Soft Gamma Repeater SGR1806-20 was observed. The main characteristics of this event are listed below:

- Peak flux: $F_{peak} > 0.3\ erg/cm^2$ (Nakar *et al.*. 2005)
- Estimated distance: $6.4\ kpc < D < 9.8\ kpc$ (Cameron *et al.*. 2005)
- Beaming angle: $\Delta\Omega \sim 0.03\ sr$ (Yamazaki *et al.*. 2005)
- γ isotropic luminosity: $1.5 \times 10^{45}\ erg \leq L_{peak}^{iso} \leq 3.5 \times 10^{45}\ erg$
- γ luminosity with beaming: $3.5 \times 10^{43}\ erg \leq L_{peak}^{beam} \leq 0.8 \times 10^{44}\ erg$

This kind of event is thought to share many of its characteristics with classical GRBs, but scaled to lower energies (Nakar *et al.*, 2005). Apart from having a different origin, they are of interest because there are other three known SGRs on the Galaxy nowadays, so that they are not of cosmological origin and might have more direct implications to life, even at present.

By applying the same method used for the GRBs before, we arrived at the $D_{10}$ distances for the different effects. The results are summarized on TABLE 3:

**TABLE 3**

As in the case of GRBs, we found that the most effective mechanism is UV Flash, because it corresponds to the largest calculated $D_{10}$ distance. However, this type of event should happen very close to Earth for dramatic effects to happen, so close that in fact we do not expect to have had any progenitors in the history of the planet.

For the case of a CRJ, we do not expect to have any direct biological effect at all given the spreading of the particles by magnetic fields (Biermann *et al.* 2004), although the acceleration of high energy particles by internal shocks in the SGR1806-20 is a possibility as discussed by Asano *et al.* (2005), it is not relevant for our purposes.

## 4. Appendix A. Population Dynamics

The use of the $D_{10}$ doses throughout this work is a way of employing a common standard of population damage; by no means should it be considered a general lethal or critical ecological threshold. For that, more complex population dynamics should be taken on account.

One approach to understand what can be considered significant population depletion, where by significant we mean that the population is on risk of being extinct, is to define the concept of Minimum Viable Population (MVP). This concept was first introduced by Shaffer (1981), but envisioning the ecological and economical problem of keeping natural reserves as small as possible, keeping the biodiversity. It is a concept, therefore, usually adopted for macroscopic populations. As Shaffer stated:

"A minimum viable population for any given species in any given habitat is the smallest isolated population having a 99% chance of remaining extant for 1000 years despite the foreseeable effects of demographic, environmental, and genetic stochasticity, and natural catastrophes."

- Shaffer itself considered it an *ad hoc* definition, because there is no special reason for the choices of 99% and 1000 years. He makes clear the necessity of adapting this definition to the system of interest.

The concept seems valid for a variety of systems: if a population becomes too small, it may end extinct. The problem is how to assess such number, which is extremely dependent on the system and its interactions with its surroundings. Shaffer proposes five ways of doing so:

- *Experiments*: the viability of doing experiments depends on the system, because it is necessary to find or create isolated populations and follow their persistence for a time scale proper of that species;

- *Biogeographic patterns*: the observations of distribution patterns that occur on islands or fragmented regions can give a first insight on the minimum areas required for the populations and, by estimating the densities, one can calculate the MVP. This approach requires the species already to be in equilibrium on the isolated regions, and that the time of isolation is known (by geological clues, for instance). Even though, the estimates may not be promptly extrapolated, because the interspecies and environmental interactions can be very distinct;

- *Theoretical models*: there are many theoretical models which can predict the probability of survival of a small population, but, these normally are not based on realistic biological hypothesis, being over-simplified, or they get into unsolved mathematical problems. The diffusion theory, as applied by many authors, can be used on totally unpredictable environments.

- *Numerical simulations*: by not suffering the limitations of the purely theoretical models, they can be the most useful way to calculate the MVP. They can be more realistic and accept many more parameters from the actual biological system, allowing their prompt modification, as well of their interdependencies. The simulations are, however, extremely specific of the modeled system, failing in giving general conclusions. They also need accurate knowledge of the critical parameters to assure a realistic simulation.

- *Genetic considerations*: many authors follow genetic and evolutionary arguments to recommend MVP. Franklin (1980) suggests that, to keep short term fitness, the effective size of the population has to be greater than 50 individuals. He also proposes that, for an environment in alteration, in order to assure sufficient genetic variability for adaptation, the number must be around 500. These recommendations are based in generalized applications of basic genetic principles, thus, they may suffer of over-simplifications.

Following the tendency to concentrate MVP calculations for macroscopic species endangered of extinction, a great deal of models takes on account the so called Allee effect, i.e., the sensibility of a population to low density of individuals due to the difficult in finding mating partners. Microorganisms, in contrast with sexually reproducing ones, are not subjected to the Allee effect. However, horizontal gene transfer (HGT) seems to be an essential mechanism on the prokaryotic domain of life (Allers and Mervarech 2005), which may imply in an analogue of the Allee effect to guarantee genetic variability. It is clear that understanding the full importance of HGT is fundamental to comprehend the evolutionary processes during the unicellular era of Earth. We must emphasize that, by neglecting the HGT, models that intend to predict the MVP for unicellular organisms may be underestimating it. As some authors suggest that up to 30% of the prokaryotic genetic material may be of HGT origin (Allers and Mervarech 2005), we must consider that the difference between a purely vertical gene transfer (VGT) model and a VGT / HGT hybrid model should be significant on the MVP calculation. Theses values are not yet present at the literature.

As Franklin (1980), other authors (Soulé 1986, Nunney and Campbell 1993) have arrived to a MVP of 10 for microorganisms. Chiba (1998) makes a systematic analytical study of the problematic of calculating the MVP in a general way, using as an example an exponentially growing population, which is a good model for microorganisms, but also neglecting HGT. On a similar approach, McCarthy (2001) calculates the MVP using stochastic methods and Monte Carlos simulations, arriving at the important result that, although populations may float abruptly with the variation of the model parameters (as perturbations), as long as the MVP is not reached, extinction should not take place. He also demonstrated that the MVP changed gradually, not abruptly, with parameters variations. This result validates the concept of MVP for extinction models.

As an experimental test for the theoretical MVP for microorganisms, Quang (1998) showed that the smallest viable population of a few species of aquatic bacteria, during the period of the experiment, fluctuated around 10 *cells/ml*.

### 5. Acknowledgements

We wish to acknowledge the partial financial support of FAPESP and CNPq Agencies (Brazil) in the form of research scholarships to the authors. The CAPES Federal Agency has funded a joint program with Santa Clara University (Cuba) from which this work has benefited.

11. Dar, A., Laor, A. and Shaviv, N.J. (1998), *Phys. Rev. Letters*, 80 (26), pp. 5813-5816

12. Dermer, C. D. and Holmes, J. M. (2005), *Astrophysical Journal Letters,* 628, pp. L21-L24

13. Ellis, J. and Schramm, D.N. (1995), *Proc. Natl. Acad. Sci USA*, 92, pp. 235-238

14. Frail, D. A., Kulkarni, S. R., Sari, R., Djorgovski, S. G., Bloom, J. S., Galama, T. J., Reichart, D. E., Berger, E., Harrison, F. A., Price, P. A., Yost, S. A., Diercks, A., Goodrich, R. W. and Chaffee, F. (2001), *Astrophysical Journal Letters*, 562, pp. L55-L58

15. Franklin, I.R. (1980), *Conservation Biology: An Evolutionary-Ecological Perspective*, Sinauer, Sunderland, MA, pp. 135-150

16. Gaisser, T. (1992), *Cosmic rays and particle physics*, Cambridge University Press

17. Gascón, J., Oubiña, A., Pérez-Lezaun, A. and Urmeneta, J. (1995), *Current Microbiology*, 30, pp. 177-182

18. Gehrels, N., Laird, C.M., Jackman, C.H., Cannizzo, J.K., Mattson, B.J. and Chen, W. (2003), *Astrophysical Journal,* 585, pp. 1169-1176

19. Ghosal, D., Omelchenko, M.V., Gaidamakova, E.K., Matrosova, V.Y., Vasilenko, A., Venkateswaran, A., Zhai, M., Kostandarithes, H.M., Brim, H., Makarova, K.S., Wackett, L.P., Fredrickson, J.K. and Daly, M.J. (2005), *FEMS Microbiology Reviews*, 29, pp. 361-375
25

**Tables**

| Test organism | $D_{10}^{ion}\,(kGy)$ | $F_{10}^{ion}\,(J/m^2)$ | $F_{10}^{UV}\,(J/m^2)$ |
|---|---|---|---|
| *E.coli* | 0.7 | $3.50\times10^5$ | 22.6 |
| *D.radiodurans* | 11 | $5.50\times10^6$ | $5.53\times10^2$ |

**Table 1**

| Planetary effect | Test Organism | $D_{10}$ distance ($kpc = 3.1\times10^{19}\,m$) | |
|---|---|---|---|
| | | *Thin atmosphere* | *Thick atmosphere* |
| γ Flash | *E.coli* | 0.39 | negligible |
| | *D.radiodurans* | 0.10 | negligible |
| UV Flash | *E.coli* | 48 - 152 | 21 - 96 |
| | *D.radiodurans* | 10 - 31 | 4 - 19 |
| Ozone Layer Depletion | *E.coli* | - | - |
| | *D.radiodurans* | - | 12 |
| CRJ | *E.coli* | - | 0.05 |
| | *D.radiodurans* | - | 0.01 |

**Table 2**

| Planetary Effect | Test organism | $D_{10}$ distance ($pc = 3.1\times10^{16}\,m$) | |
|---|---|---|---|
| | | *Thin atmosphere* | *Thick atmosphere* |
| γ Flash | *E.coli* | 0.06 – 0.09 | negligible |
| | *D.radiodurans* | 0.01 – 0.02 | negligible |
| UV Flash | *E.coli* | 7 - 35 | 3 - 22 |
| | *D.radiodurans* | 1 - 7 | 1 - 4 |
| Ozone Layer Depletion | *E.coli* | - | - |
| | *D.radiodurans* | - | 2 - 3 |
| CRJ | *E.coli* | - | 0.01 |
| | *D.radiodurans* | - | negligible |

**Table 3**

**Table legends**

**Table 1**

Doses and fluxes for a 10% survival for the bacteria *E.coli* and *D.radiodurans* for ionizing (Ghosal et al. 2005) and UV radiation (Gascón et al. 1995).

**Table 2**

$D_{10}$ distances for the GRB effects, clearly showing the longer range for the UV Flash mechanism. The quoted ranges for the UV Flash reflect the uncertainty in the fraction of γ to UV conversion efficiency (see text). Nor the ozone depletion nor the CRJ effects is relevant on the thin atmosphere case, and for *E.coli*, the $D_{10}$ distance is not calculated for ozone depletion because it is already bellow the $D_{10}$ threshold without the burst.

**Table 3**

$D_{10}$ distances for the SGR effects. The quoted ranges reflect the uncertainty on the distance to the source and on the fraction of γ to UV conversion efficiency (see text). Nor the ozone depletion nor the CRJ effects is relevant on the thin atmosphere case, and for *E.coli*, the $D_{10}$ distance is not calculated for ozone depletion because the solar flux on the ground is already above the $D_{10}$ threshold without ozone loss.

**Figures**

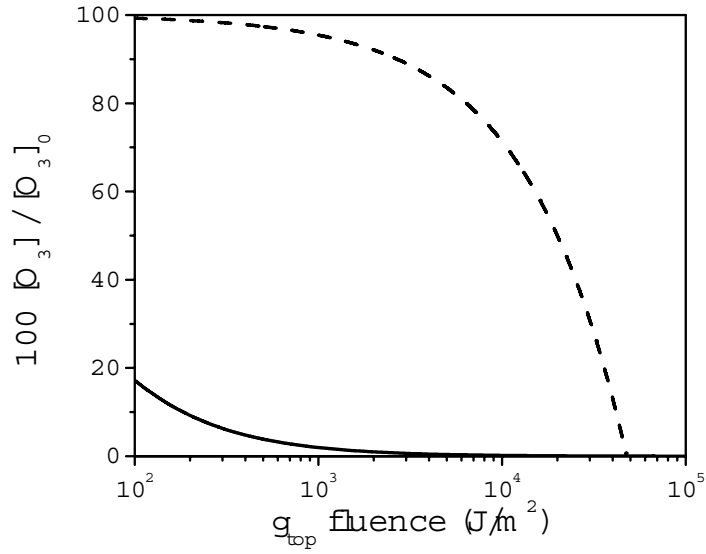

**Fig 1**

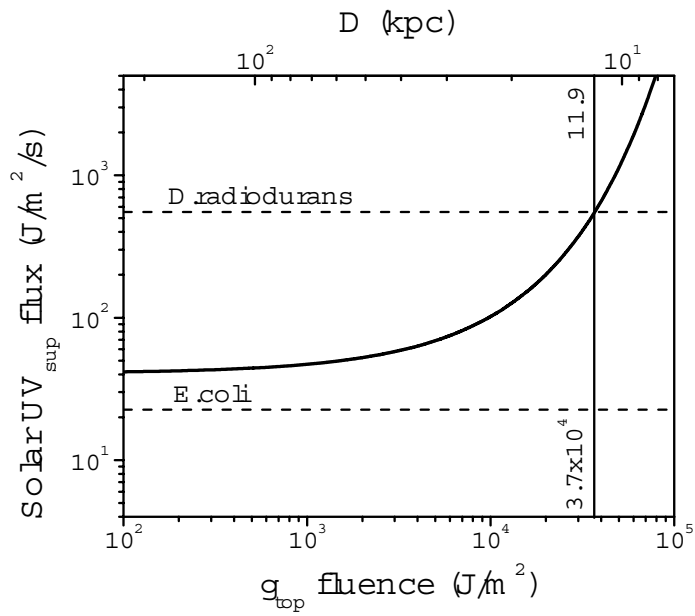

**Fig 2**

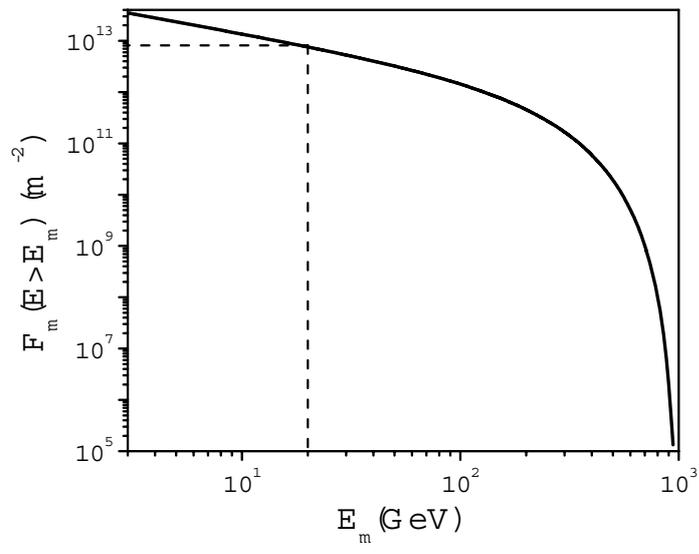

**Fig 3**

**Figure legends**

**Fig 1**

Atmospheric ozone depletion as a function of the initial γ fluence, as calculated by Ruderman (1979) (solid line), using an analytical simplified approach, and by Gehrels et al. (2003) (dashed line), using numerical simulations.

**Fig 2**

Calculated solar UV flux reaching sea level, attenuated by the depleted ozone layer, as a function of the initial $\gamma_{top}$ fluence (lower scale) and distance to the standard GRB (upper scale). The $F_{10}^{UV}$ for E.coli and D.radiodurans are also plotted.

**Fig 3**

Muon flux on sea level produced by the interactions of 1 TeV protons on top of the atmosphere, not considering muon decay. For muons over 20 GeV, the expected muon flux is of the order of $10^{13}$ m$^{-2}$.